\documentclass[aps,prl,reprint,showpacs,superscriptaddress,longbibliography]{revtex4-2}
\usepackage{amsmath,amssymb,graphicx}
\usepackage{graphicx,epstopdf}
\usepackage{gensymb}
\usepackage[dvipsnames]{xcolor}
\usepackage[T1]{fontenc}
\usepackage[utf8]{inputenc}
\newcommand{\be}{\begin{equation}}
\newcommand{\ee}{\end{equation}}
\newcommand{\bea}{\begin{eqnarray}}
\newcommand{\eea}{\end{eqnarray}}
\newcommand{\bse}{\begin{subequations}}
\newcommand{\ese}{\end{subequations}}

\usepackage{multibib}
\usepackage{color}
\usepackage[colorlinks,bookmarks=false,citecolor=darkblue,linkcolor=red,urlcolor=blue]{hyperref}

\definecolor{darkred}{rgb}{0.7,0.0,0.0}

\definecolor{darkblue}{rgb}{0,0.02,0.45}
\def\cdbl{\color{darkblue}}
\definecolor{darkgreen}{rgb}{0.02,0.45,0.0}

\definecolor{violet}{rgb}{0.8,0.2,0.6}

\begin{document}
	
\title{Quantum disordered ground state in the spin-orbit coupled $J_{\rm eff}$ = $\frac{1}{2}$ distorted honeycomb magnet BiYbGeO$_{5}$}

\author{S. Mohanty}
\author{S. S. Islam}
\affiliation{School of Physics, Indian Institute of Science Education and Research Thiruvananthapuram-695551, India}

\author{N. Winterhalter-Stocker}
\author{A. Jesche}
\affiliation{Center for Electronic Correlations and Magnetisim, University of Augsburg, 86159 Augsburg, Germany}

\author{G. Simutis}
\affiliation{Laboratory for Neutron and Muon Instrumentation, Paul Scherrer Institut, CH-5232 Villigen PSI, Switzerland}

\author{Ch. Wang}
\author{Z. Guguchia}
\affiliation{Laboratory for Muon Spin Spectroscopy, Paul Scherrer Institut, Villigen PSI, Switzerland}

\author{J. Sichelschmidt}
\author{M. Baenitz}
\affiliation{Max Planck Institute for Chemical Physics of Solids, Nothnitzer Str: 40, 01187 Dresden, Germany}

\author{A. A. Tsirlin}
\affiliation{Felix Bloch Institute for Solid-State Physics, Leipzig University, 04103 Leipzig, Germany}

\author{P. Gegenwart}
\affiliation{Center for Electronic Correlations and Magnetisim, University of Augsburg, 86159 Augsburg, Germany}

\author{R. Nath}
\email{rnath@iisertvm.ac.in}
\affiliation{School of Physics, Indian Institute of Science Education and Research Thiruvananthapuram-695551, India}
\date{\today} 

\begin{abstract}
We delineate quantum magnetism in the strongly spin-orbit coupled, distorted honeycomb-lattice antiferromagnet BiYbGeO$_{5}$.
Our magnetization and heat capacity measurements reveal that its low-temperature behavior is well described by an effective $J_{\rm eff}=1/2$ Kramers doublet of Yb$^{3+}$. The ground state is nonmagnetic with a tiny spin gap.
Temperature-dependent magnetic susceptibility, magnetization isotherm, and heat capacity could be modeled well assuming isolated spin dimers with anisotropic exchange interactions $J_{\rm Z} \simeq 2.6$~K and $J_{\rm XY} \simeq 1.3$~K. 
Heat capacity measurements backed by muon spin relaxation suggest the absence of magnetic long-range order down to at least 80\,mK both in zero field and in applied fields. This sets BiYbGeO$_5$ apart from Yb$_2$Si$_2$O$_7$ with its unusual regime of magnon Bose-Einstein condensation and suggests negligible interdimer couplings, despite only a weak structural deformation of the honeycomb lattice.
\end{abstract}

\maketitle

{\cdbl\textit{Introduction.}} 
Antiferromagnetic spin-$1/2$ dimer is the simplest case of a quantum magnet
characterized by the singlet ($S = 0$) ground state with entangled spins and an excitation gap in the energy spectrum. 
Closing this gap by applying external magnetic fields has been instrumental in stabilizing long-range order in spin-dimer systems~\cite{Zapf563}. For the SU(2) symmetry of the underlying spin Hamiltonian, such an order is often described in terms of Bose-Einstein condensation of magnons~\cite{Rice760,Giamarchi198}. Experimental manifestations of this scenario include field-induced ordered states of $3d$ magnets with the spin-dimer geometry~\cite{Nikuni5868,Zapf563,Giamarchi198,Jaime087203,Mukharjee144433}. More recently, similar effects were observed in the $4f$ magnet Yb$_2$Si$_2$O$_7$, although two distinct ordered states have been reported in this case~\cite{Hester027201}. Different microscopic mechanisms were proposed for this behavior, including the weak, hitherto not detected anisotropy of Yb$^{3+}$~\cite{Flynn067201} and the special geometry of interdimer interactions that arises from the underlying honeycomb lattice~\cite{Feng00032}. 



Magnetism of the Yb$^{3+}$ ions is often anisotropic. Anisotropic nature of the ion itself, as well as exchange anisotropy of magnetic couplings may be behind many interesting effects, including the possible realization of quantum spin ice in pyrochlore materials~\cite{Ross021002,Thompson2011,Thompson2017}, persistent dynamics observed in triangular spin-liquid candidates~\cite{Li2020,Paddison2017}, and Tomonaga-Luttinger liquid behavior with spinon confinement-deconfinement transitions in spin-chain magnets~\cite{Wu698}. Yb-based honeycomb magnets are currently studied as potential hosts for the Kitaev physics~\cite{Xing014427,Sala2021,Zhang033006,Wessler2020}. Concurrently, deformed honeycomb lattices of Yb$^{3+}$ could set an interesting link to Yb$_2$Si$_2$O$_7$ and reveal the effect of magnetic anisotropy on the field-induced states of a dimer magnet.

\begin{figure}
	\includegraphics[width= \linewidth]{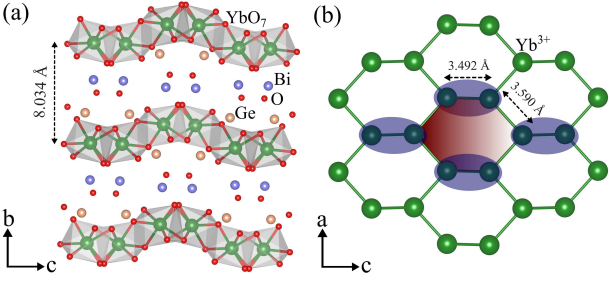}
	\caption{\label{Fig1} (a) Crystal structure of BiYbGeO$_5$ viewed along the $a$-axis, showing well separated honeycomb layers formed by edge-shared YbO$_7$ polyhedra. (b) A section of the honeycomb layer projected onto the $ac$-plane highlights the short \mbox{Yb$^{3+}$--Yb$^{3+}$} distance that leads to spin dimerization.}
\end{figure}

Herein, we report one such case, the quantum magnet BiYbGeO$_5$~\cite{Cascales3626} that features a quasi-2D distorted honeycomb lattice of Yb$^{3+}$ ions in the $ac$-plane of the structure [see Fig.~\ref{Fig1}]. This geometry is very similar to the one known from Yb$_2$Si$_2$O$_7$~\cite{Hester027201}. In the BiYbGeO$_5$ case, the 2D honeycomb layers are strongly buckled and, consequently, deformed. On the other hand, the two nearest-neighbor Yb$^{3+}$ -- Yb$^{3+}$ distances remain similar, $d_1=3.492$~\AA\ (dimer bond, $J_0$) and $d_2=3.590$~\AA\ (interdimer bonds, $J'$), respectively. We elucidate the low-energy states of Yb$^{3+}$ as Kramers doublet described by an effective spin $J_{\rm eff}=1/2$, establish the dimerized regime ($J_0>J'$) with quantum disordered ground state, and examine possibility of the field-induced magnetic order in this material.

{\cdbl\textit{Experimental.}} 
Polycrystalline sample of BiYbGeO$_5$ was synthesized via the conventional solid state synthesis technique. The phase purity of the compound was confirmed from the powder x-ray diffraction (XRD) and the subsequent Rietveld refinement of the XRD data [see the Supplementary Material (SM)~\cite{supplementary}]. Magnetization ($M$) was measured with the help of a SQUID magnetometer (MPMS-3, Quantum Design) down to 0.4~K with the $^3$He (iHelium3, Quantum Design, Japan) attachment. Heat capacity ($C_{\rm p}$) was measured on a piece of pellet using the heat capacity option in the PPMS (Quantum Design). The data down to 80\,mK were determined by the relaxation technique in a dilution refrigerator cryostat.

Muon spin relaxation ($\mu^{+}$SR) measurement was performed at the S$\mu$S muon source at Paul Scherrer Institute using a combination of two spectrometers (GPS and HAL) down to below 100~mK in zero-field. The details of the $\mu^{+}$SR experiment is described in the SM~\cite{supplementary}.

Magnetization and heat capacity of the anisotropic spin dimer were obtained by exact diagonalization using the \texttt{fulldiag} utility of the \verb"ALPS" package~\cite{BauerP05001}.


\begin{figure}
	\includegraphics[width=\columnwidth]{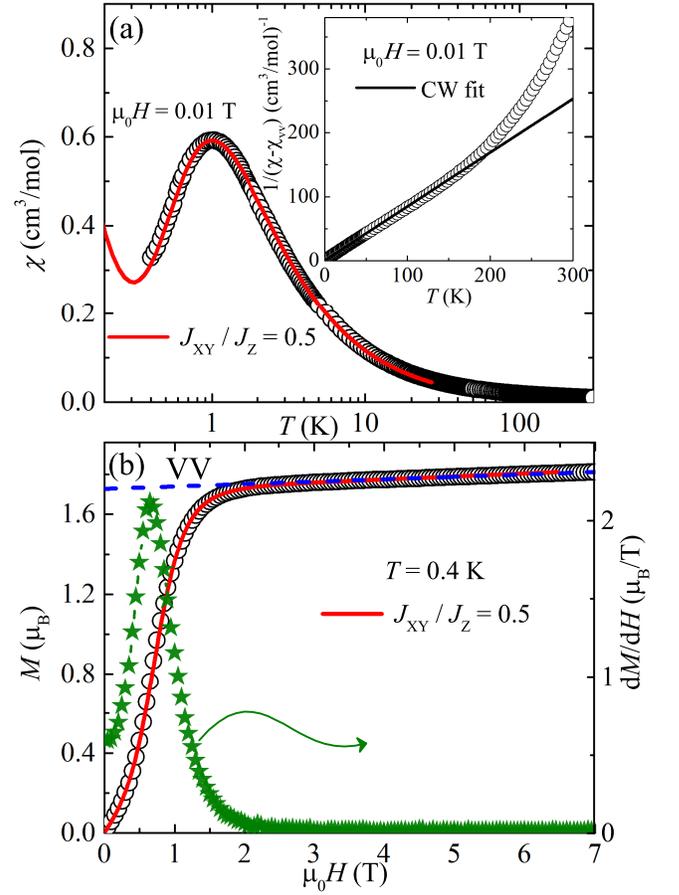}
	\caption{\label{Fig2} (a) $\chi(T)$ of BiYbGeO$_5$ measured in $\mu_0 H = 0.01$~T. The solid line represents the simulation for isolated spin dimers with anisotropic interactions ($J_{\rm z} = 2.6$~K and $J_{\rm xy} = 1.3$~K). Inset: CW fit to the low-$T$ $1/\chi$ data (after subtracting the Van-Vleck contribution). (b) $M$ vs $H$ and $dM/dH$ vs $H$ in the left and right $y$-axes, respectively measured at $T = 0.4$~K. The horizontal dashed line marks the Van-Vleck contribution. The solid line shows the simulation.}
\end{figure}

{\cdbl\textit{Magnetization.}} 
Magnetic susceptibility ($\chi$) as a function of $T$ in an applied field $\mu_{0}H \simeq 0.01$~T is depicted in Fig.~\ref{Fig2}(a). $\chi(T)$ increases with decreasing $T$ in a Curie-Weiss (CW) manner and portrays a broad maximum centered at around 1~K, followed by a rapid decrease. No clear signature of magnetic long-range-order (LRO) is evident down to 0.4~K. A weak upturn at very low temperatures is likely due to the presence of a small extrinsic paramagnetic contribution~\cite{Tsirlin104436}. The broad maximum reflects the short-range correlations anticipated for a low-dimensional AFM spin system and the rapid decrease signals the opening of a spin gap at low temperatures~\cite{Arjun014421,Tsirlin144412,Mukharjee224403}.

Above 150~K, the inverse susceptibility ($1/\chi$) for $H = 0.01$~T was fitted well by $\chi(T)=\chi_0+\frac{C}{T-\theta_{\rm CW}}$, where $\chi_0$  is the temperature-independent susceptibility and the second term is the CW law. The fit yields $\chi_0 \simeq 1.5 \times 10^{-3}$~cm$^3$/mol, the high-$T$ effective moment $\mu_{\rm eff}^{\rm HT}$ $[=\sqrt{3k_{\rm B}C/N_{\rm A}}$, where $C$, $k_{\rm B}$, and $N_{\rm A}$ are the Curie constant, Boltzmann constant, and Avogadro’s number, respectively]~$\simeq 4.78~\mu_{\rm B}$, and the high-$T$ CW temperature $\theta_{\rm CW}^{\rm HT} \simeq -67.2$~K (see the SM~\cite{supplementary}). This value of $\mu_{\rm eff}^{\rm HT}$ is in good agreement with the expected value, $\mu_{\rm eff} = g\sqrt{J(J+1)}~\simeq 4.54~\mu_{\rm B}$ for Yb$^{3+}$ ($J = 7/2$, $g=8/7$) in the $^{4}f_{13}$ configuration.

\begin{figure*}
	\includegraphics[width=\textwidth]{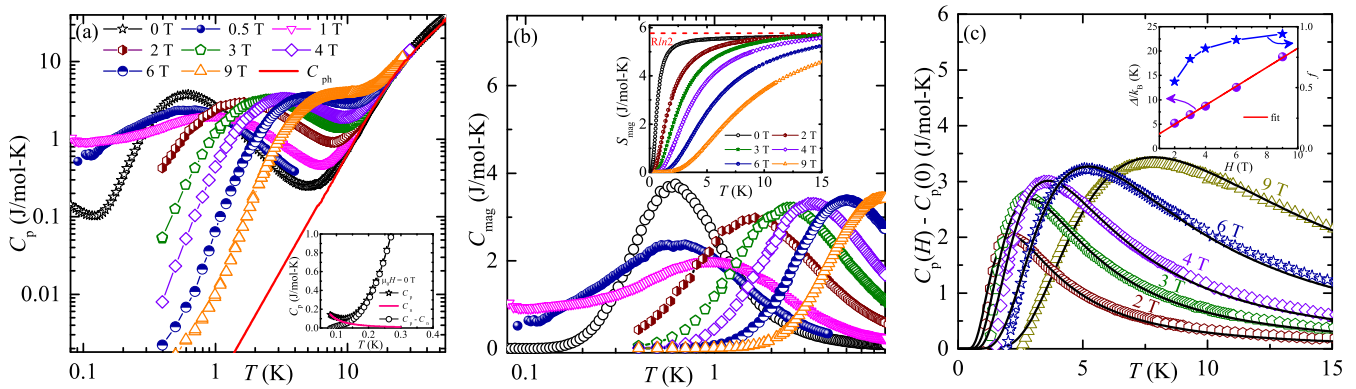}
	\caption{\label{Fig3} (a) $C_{\rm p}$ vs $T$ for BiYbGeO$_5$ measured in different applied fields. The solid line represents $C_{\rm ph}(T)$ of the nonmagnetic analog BiYGeO$_5$. Inset: Zero-field $C_{\rm p}$ vs $T$ in the low temperature regime. The solid line represents the nuclear contribution ($C_{\rm n}$). (b) Magnetic heat capacity $C_{\rm mag}$ vs $T$ in different magnetic fields. Inset: Magnetic entropy $S_{\rm mag}$ vs $T$ for different magnetic fields. (c) Electronic Schottky contribution [$C_{\rm p}(H)-C_{\rm p}(0)$] vs $T$ along with the fit using Eq.~\eqref{Eqelectronicschottky}. Inset: $\Delta/k_{\rm B}$ and $f$ vs $H$ in the left and right $y$-axes, respectively. The solid line represents the straight line fit to $\Delta/k_{\rm B}$ vs $H$.}
\end{figure*}

Inverse susceptibility, $1/\chi$, is found to deviate from linearity at low temperatures with a clear slope change. After subtraction of the Van-Vleck susceptibility ($\chi_{\rm VV}$), obtained from the $M$ vs $H$ analysis, $1/(\chi-\chi_{\rm VV})$ shows a linear regime at low temperatures [inset of Fig.~\ref{Fig2}(a)]. A CW fit in the $T$-range of $10-30$~K yields $\mu_{\rm eff} \simeq 3.07~\mu_{\rm B}$ and $\theta_{\rm CW} \simeq -0.67$~K. The negative value of $\theta_{\rm CW}$ reflects dominant AFM exchange between the Yb$^{3+}$ ions. This experimental $\mu_{\rm eff}$ corresponds to an effective spin $J_{\rm eff}=1/2$ with an average $g\simeq 3.5$~\cite{Somesh064421}. Such a large value of $g$ compared to the free-electron value of $g = 2.0$ reflects strong spin-orbit coupling and is consistent with the one obtained from the ESR experiments (see SM)~\cite{supplementary}. In Yb$^{3+}$-based compounds, the Kramers doublets ($m_{J} = \pm 1/2$) evoked by the CEF excitations essentially control the magnetic properties at low temperatures. In such a scenario, the ground state is an effective $J_{\rm eff}=1/2$ state, while the excited states generate a significant $\chi_{\rm VV}$~\cite{Li035107,Li167203,Ranjith115804}.
Our $\chi(T)$ analysis supports the interpretation in terms of an effective pseudo-spin-1/2 ground state, similar to other Yb$^{3+}$-based compounds~\cite{Bag220403,*Schmidt214445,Li035107}.

Figure~\ref{Fig2}(b) presents the magnetic isotherm ($M$ vs $H$) measured at $T = 0.4$~K. It manifests a distinct curvature and then the tendency of saturation but increases very weakly with further increase in field due to Van-Vleck contribution. The linear fit to the high-field ($\geq6$~T) data returns a slope of $\sim 0.012 \mu_{\rm B}/{\rm T}$, which corresponds to $\chi_{\rm VV}\simeq 6.7 \times 10^{-3}$~cm$^{3}$/mol. From the $y$-intercept of the linear fit, the saturation magnetization is estimated to be $M_S \simeq 1.7~\mu_{\rm B}$, which is in good agreement with $M_{\rm S} = gJ_{\rm eff}\mu_{\rm B} \simeq 1.75~\mu_{\rm B}$, expected for $J_{\rm eff}=1/2$ with powder-averaged $g\simeq3.5$~\cite{Somesh064421}. Thus, the $M$ vs $H$ analysis also supports the $J_{\rm eff}=1/2$ ground state of Yb$^{3+}$.

{\cdbl\textit{Heat capacity.}} 
Zero-field heat capacity [Fig.~\ref{Fig3}(a)] exhibits a broad maximum at $\sim 0.6$~K and then decreases rapidly with temperature. While the broad maximum is attributed to the onset of short-range correlations, the rapid decrease is a hallmark of the formation of the singlet ground state~\cite{Zapf563,Freitas184426}.
A small upturn at very low temperatures ($< 0.1$~K) may be ascribed to the nuclear contribution to $C_{\rm p}$~\cite{Freitas184426}. When magnetic field is applied, initially the broad maximum is strongly suppressed for $\mu_0H < 1$~T suggesting the suppression of AFM correlations and the closing of spin gap~\cite{Ruegg247202}. For $H > 1$~T, the position of the broad maximum shifts toward high temperatures and its amplitude is enhanced significantly. Finally, for $\mu_0H > 2$~T the height of the maximum almost saturates in the fully polarized state. This magnetic field-driven broad maximum is a clear testimony of the Schottky anomaly due to the Zeeman splitting of the ground state Kramers doublet.

Heat capacity of the nonmagnetic analog BiYGeO$_5$, which represents the phononic contribution ($C_{\rm ph}$), was measured down to 2~K and extrapolated to 80~mK by a $T^{3}$ fit to the low-temperature data. Magnetic heat capacity ($C_{\rm mag}$) obtained by subtracting $C_{\rm ph}$ from the measured $C_{\rm p}$ in different fields is shown in Fig.~\ref{Fig3}(b).
In zero field, the nuclear contribution is removed by fitting the data below 0.12~K by $C_{\rm n}(T) = \alpha_{\rm Q}/T^{2}$ [see inset of Fig.~\ref{Fig3}(a)]. The fit yields the coefficient $\alpha_{\rm Q} \simeq 9.1 \times 10^{-4}$~J~K/mol, which is in reasonable agreement with that reported for other Yb-based systems~\cite{Ranjith224417}. The obtained $C_{\rm mag}(T)$ is then used to estimate the magnetic entropy [$S_{\rm mag}(T)$] by integrating $C_{\rm mag}(T)/T$ in the measured $T$-range [inset of Fig.~\ref{Fig3}(b)]. $S_{\rm mag}(T)$ reaches a well-defined plateau at $R\ln2$ in zero field, further endorsing the fact that the low-temperature properties can be explained by the $J_{\rm eff}=1/2$ state~\cite{Li167203,Guo094404,Tokiwa42}. In zero field, $C_{\rm mag}$ decays rapidly below the broad maximum, reflecting the singlet ground state.

To evaluate the Schottky contribution quantitatively, the zero-field data $C_{\rm p}(T,H=0)$ are subtracted from the high-field data $C_{\rm p}(T,H)$ [i.e. $C_{\rm Sch}(T,H) = C_{\rm p}(T,H) - C_{\rm p}(T,H = 0)$]. In Fig.~\ref{Fig3}(c), $C_{\rm Sch}(T,H)$ is fitted by the two-level Schottky function~\cite{Kittelc2005}
\begin{equation}
	C_{\rm Sch}(T) = fR\left(\frac{\Delta}{k_{\rm B}T}\right)^{2} \frac{e^{(\Delta/k_{\rm B}T)}}{\left[1+e^{(\Delta/k_{\rm B}T)}\right]^2},
	\label{Eqelectronicschottky}
\end{equation}
where $f$ is the fraction of free spins excited by the applied field, $\Delta/k_{\rm B}$ is the crystal-field gap between the ground state and the first excited Kramers doublet, and $R$ is the gas constant. The inset of Fig.~\ref{Fig3}(c) presents the obtained $f$ and $\Delta/k_{\rm B}$ as a function of $H$ in the right and left $y$-axes, respectively. $f$ increases with $H$ and then attains a constant value of about $\sim 1$ for $\mu_0H>3$~T, confirming that magnetic field splits the energy levels, excites the free Yb$^{3+}$ spins to the higher-energy levels, and nearly 100\% spins are excited above the saturation field. 
Similarly, the maximum of the $C_{\rm p}(H) - C_{\rm p}(0)$ curves almost attains a constant value for $\mu_0H>3$~T, suggesting that $\sim 100$\% spins become free in higher fields. Further, $\Delta/k_{\rm B}$ increases linearly with $H$ and a straight line fit returns the zero-field energy gap $\Delta/k_{\rm B}(0) \simeq 1.07$~K that possibly indicates an intrinsic field in the system~\cite{Kundu117206}. Using the value of $\Delta/k_{\rm B} \simeq 18.8$~K at 9~T in $\Delta/k_B = g\mu_{\rm B}H/k_{\rm B}$, the $g$-value is estimated to be $g \simeq 3.5$, which is consistent with the magnetization analysis.

\begin{figure}
	\includegraphics[width=\columnwidth]{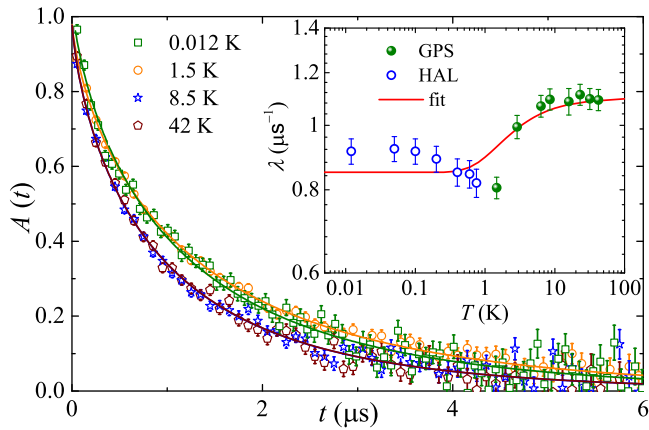}
	\caption{\label{Fig4} (a) Muon decay asymmetry vs time in zero field at four different temperatures with solid lines being the exponential fits. Inset: $\lambda$ vs $T$ together with the fit using an activated behaviour (solid line) with $\Delta_{\mu}/k_{\rm B} \simeq 1.7$~K.}
\end{figure}


{\cdbl\textit{Muon spin relaxation.}} 
The muon asymmetry curves measured in zero field and in different temperatures are displayed in Fig.~\ref{Fig4}. No oscillations were resolved in the muon spin polarization down to 12~mK, corroborating the absence of magnetic LRO.
The asymmetry curves follow a stretched exponential behavior $A(t) = A(0)\,e^{-(\lambda t)^\beta}$ with only little temperature dependence, a footprint of the dynamics of a disordered (singlet) state~\cite{Yaouanc2004}.
The stretched exponent $\beta$ is found to be $\sim 0.73$ at all temperatures. The estimated depolarization rate ($\lambda$) as a function of $T$ is presented in the inset of Fig.~\ref{Fig4}. It is almost constant at high temperatures, exhibits a drop below around $\sim 4$~K, and reaches a constant value below around 1~K. This behavior is reproduced well by an exponential function $\lambda \propto e^{-(\Delta_{\mu}/k_{\rm B}T)}$ with a spin gap $\Delta_{\mu}/k_{\rm B} \simeq 1.7$~K, further endorsing a singlet ground state with no magnetic LRO~\cite{Williams013082}.


{\cdbl\textit{Discussion.}} 
Deformation of the honeycomb spin lattice in BiYbGeO$_5$ potentially allows different microscopic regimes: i) weakly distorted honeycombs ($J_0\simeq J'$); ii) spin dimers ($J_0\gg J'$); and iii) spin chains ($J'\gg J_0$). The first scenario should lead to a long-range order already in zero field, as previously observed in YbCl$_3$ with its weakly distorted honeycombs~\cite{Xing014427,Sala2021}. The absence of long-range order in zero field excludes this scenario. Assuming Heisenberg interactions, spin dimer can be distinguished from the spin chain by the presence or absence of a spin gap, respectively. The gapped state observed experimentally in BiYbGeO$_5$ favors the dimer scenario.

\begin{figure}
	\includegraphics[width=\columnwidth]{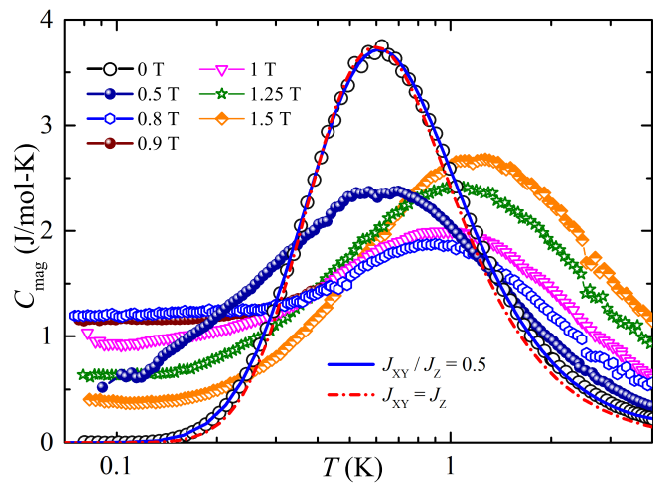}
	\caption{\label{Fig5} Magnetic heat capacity, $C_{\rm mag}$ vs $T$, measured down to 0.08~K in different fields. The solid and dashed lines represent simulated zero-field curves for an isolated spin-$1/2$ dimer with either isotropic or anisotropic exchange interactions.}
\end{figure}

Direct simulation of the magnetic heat capacity for the Heisenberg spin dimer shows a reasonable agreement with the experiment (Fig.~\ref{Fig5}). However, a close inspection of the data reveals that the maximum of the simulated curve is more narrow than the experimental one. This discrepancy can be remedied by considering an anisotropic exchange coupling,
\begin{equation}
 \mathcal H=J_{\rm XY}(S_i^xS_j^x+S_i^yS_j^y)+J_{\rm Z}S_i^zS_j^z,
\label{eq:ham}\end{equation}
with $J_{\rm XY} \simeq 1.3$~K and $J_{\rm Z} \simeq 2.6$~K. The anisotropy $J_{\rm XY}/J_{\rm Z}\simeq 0.5$ controls the width of the maximum and can be determined rather accurately even with powder data. Temperature-dependent susceptibility and field-dependent magnetization are well reproduced with the same parameters assuming isotropic $g=3.5$ (see Fig.~\ref{Fig2}). The van Vleck term, $\chi_{\rm VV}=0.012$\,$\mu_B$/T, was added in the case of $M(H)$, whereas the susceptibility fit included an impurity contribution $C_{\rm imp}/(T-\theta_{\rm imp})$ with $C_{\rm imp}=0.05$~cm$^3$\,K/mol and $\theta_{\rm imp}=0.07$~K.

The excellent fit of thermodynamic data with the model of isolated spin dimers, Eq.~\eqref{eq:ham}, suggests that any interdimer couplings must be negligible ($\leq 60$~mK). The weakness of these couplings would be in line with the fact that signatures of long-range magnetic order have been observed neither in zero field nor in applied fields down to at least 80\,mK. We thus conclude that BiYbGeO$_5$ is well described by the model of isolated anisotropic spin dimers. This result may look unexpected considering a rather weak geometrical distortion of the honeycomb layer, $2(d_1-d_2)/(d_1+d_2)=2.8\%$, which is comparable to 3.4\% in Yb$_2$Si$_2$O$_7$~\cite{Hester027201}. However, the honeycomb layers in BiYbGeO$_5$ are strongly buckled, unlike in the silicate. According to Fig.~\ref{Fig1}, the dimer coupling $J_0$ occurs in the flat part of the layer, whereas the interdimer bonds $J'$ lie on the folding line and may be strongly affected by the buckling. Moreover, the coordination of Yb$^{3+}$ changes from 6-fold in Yb$_2$Si$_2$O$_7$ to 7-fold in BiYbGeO$_5$. This may change the nature of the ground-state Kramers doublet and lead to a significant modification of the exchange couplings.

{\cdbl\textit{Summary.}} 
BiYbGeO$_5$ features a distorted honeycomb arrangement of the $J_{\rm eff}=1/2$ Yb$^{3+}$ ions. We have shown that magnetism of this material is well described by the model of anisotropic spin dimers. The quantum disordered state with a spin gap is observed in zero magnetic field, and no LRO redolent of Bose-Einstein condensation of triplons is induced by the applied field down to at least 80~mK, contrary to the typical spin-dimer magnets. We ascribe this behavior to the fact that spin dimers are magnetically nearly isolated. This sets BiYbGeO$_5$ apart from another Yb-based dimer magnet, Yb$_2$Si$_2$O$_7$, that shows a similar deformation of the honeycomb lattice but without the buckling of the structural layers.



{\cdbl\textit{Acknowledgments.}} 
We would like to acknowledge SERB, India for financial support bearing sanction Grant No. CRG/2022/000997. Part of this work is based on experiments performed at the Swiss Muon Source S$\mu$S, Paul Scherrer Institute, Villigen, Switzerland. The project has received funding from the European Union’s Horizon 2020 research and innovation programme under the Marie Sk{\l}odowska-Curie grant agreement No. 884104 (PSI-FELLOW-III-3i). Computations for this work were done using resources of the Leipzig University Computing Center. This work was supported by the German
Research Foundation (DFG) via Project No. 107745057 (TRR80).

%

\end{document}